# Ribonucleocapsid assembly/packaging signals in the genomes of the coronaviruses SARS-CoV and SARS-CoV-2: Detection, comparison and implications for therapeutic targeting

Vladimir R. Chechetkin[a]* and Vasily V. Lobzin[b]

[a]*Engelhardt Institute of Molecular Biology of Russian Academy of Sciences, Vavilov str., 32, Moscow, Russia*

[b]*School of Physics, University of Sydney, Sydney, NSW 2006, Australia*

___________________________

*Corresponding author. *E-mail addresses:* chechet@eimb.ru; vladimir_chechet@mail.ru

Tel.: +7 499 135 9895. Fax: +7 499 135 1405 (V.R. Chechetkin).

**Abstract**

The genomic ssRNA of coronaviruses is packaged within a helical nucleocapsid. Due to transitional symmetry of a helix, weakly specific cooperative interaction between ssRNA and nucleocapsid proteins leads to the natural selection of specific quasi-periodic assembly/packaging signals in the related genomic sequence. Such signals coordinated with the nucleocapsid helical structure were detected and reconstructed in the genomes of the coronaviruses SARS-CoV and SARS-CoV-2. The main period of the signals for both viruses was about 54 nt, that implies 6.75 nt per N protein. The complete coverage of ssRNA genome of length about 30,000 nt by the nucleocapsid would need 4,400 N proteins, that makes them the most abundant among the structural proteins. The repertoires of motifs for SARS-CoV and SARS-CoV-2 were divergent but nearly coincided for different isolates of SARS-CoV-2. We obtained the distributions of assembly/packaging signals over the genomes with non-overlapping windows of width 432 nt. Finally, using the spectral entropy, we compared the load from point mutations and indels during virus age for SARS-CoV and SARS-CoV-2. We found the higher mutational load on SARS-CoV. In this sense, SARS-CoV-2 can be treated as a "newborn" virus. These observations may be helpful in practical medical applications and are of basic interest.

**Keywords**

COVID-19, SARS-CoV, SARS-CoV-2, Helical nucleocapsid, Genome packaging.

**List of Abbreviations**

DDFT, discrete double Fourier transform; DFT, discrete Fourier transform; NCF, nucleotide correlation functions; N proteins, nucleocapsid proteins; SARS-CoV, severe acute respiratory syndrome coronavirus; SARS-CoV-2, severe acute respiratory syndrome coronavirus 2; ssRNA, single-stranded RNA; TAMGI, transitional automorphic mapping of the genome on itself; UTR, untranslated region

# 1. Introduction

To the end of July 2020, the COVID-19 pandemia was the cause of more than 17.8 millions of coronavirus cases and more than 680,000 of deaths over all world (https://www.worldometers.info/coronavirus/). The pandemia is still continuing and the possibility of return of new disease waves is considered to be very high. The development of efficient medications and vaccines against coronaviruses needs the knowledge of main molecular mechanisms in the virus life cycle and virus-host interaction (Maier et al., 2015; Ziebuhr, 2016; Fung & Liu, 2019; Saxena, 2020; Feng et al., 2020; Xie & Chen, 2020). In this paper we will discuss a specific interaction between nucleocapsid (N) proteins and genomic ssRNA in the coronaviruses SARS-CoV and SARS-CoV-2.

The ssRNA genome of the coronaviruses is packaged within a helical nucleocapsid, while the whole ribonucleocapsid is packaged within a membrane envelope (for a review see, e.g., Neuman & Buchmeier, 2016; Masters, 2019). The term "packaging signal" in the coronavirus papers is overwhelmingly attributed to the specific interaction between genomic RNA and membrane (M) proteins ensuring the transport of the ribonucleocapsid into the membrane envelope (Fosmire et al., 1992; Narayanan, & Makino, 2001; Madhugiri et al., 2016; Woo et al., 2019; Masters, 2019). The interactions between the genomic RNA and N proteins are assumed to be non-specific and governed mainly by electrostatic effects. The question of how N proteins recognize the relative genomic RNA remains unanswered. A similar point of view was long-lastingly prevalent also in the virology community which studied ssRNA viruses with icosahedral capsids. The importance of cooperative weakly specific interactions between ssRNA and capsid proteins has been recognized not long ago (see discussion by Twarock et al. (2017) and references therein). Stockley et al. (2016) suggested and proved experimentally a two-stage model for the assembly of ssRNA viruses with icosahedral capsids. At the first, more rapid, stage RNA binds to the coat proteins to facilitate capsid assembly, whereas at the second, slower, stage RNA is compactly packaged within the capsid. The specific cooperative RNA-coat protein interactions play important role at the both stages. The two stages may be associated with different signals (Chechetkin & Lobzin, 2019) and the whole dynamic process may be called assembly/packaging. The generalization of these ideas on viruses with ribonucleocapsid within the membrane envelope like that for coronaviruses assumes three stages related to the complete packaging of the genomic RNA within envelope: two-staged assembly/packaging of the helical ribonucleocapsid and packaging of the ribonucleocapsid within the envelope. This paper is devoted to search for specific signals in the genomic ssRNA sequences related to two-staged assembly/packaging of the helical ribonucleocapsid. As has been shown previously, the icosahedral symmetry of the capsid strongly affects the large-scale quasi-periodic segmentation in the related viral genomes (Chechetkin & Lobzin, 2019). The whole ribonucleocapsid structure of coronaviruses also remains invariant under transition by one helical turn. Therefore, the putative weakly specific assembly/packaging signals in the genomic RNA of coronaviruses should be coordinated with the parameters of the helical nucleocapsid (such as the helix pitch, inner and outer diameters) which are established by cryo-electron microscopy (cryo-EM) and other structural methods. In this paper we provide methods for the detection and comparative analysis of assembly/packaging signals in the genomic RNA of the coronaviruses SARS-CoV and SARS-CoV-2 and describe main results of our study.

## 2. Theory and methods

The quasi-periodic patterns in the genomic DNA/RNA sequences can be efficiently detected with the discrete Fourier transform (DFT). As the periodic patterns generate equidistant series of harmonics in the DFT spectra (see, e.g., Chechetkin & Turygin, 1995; Lobzin & Chechetkin, 2000), long enough patterns can be detected by the iteration of DFT or by the discrete double Fourier transform (DDFT) (Chechetkin & Lobzin, 2017, 2019, 2020a, b). Though the correlation functions are the main tools in this paper, our approach is based implicitly and explicitly on DFT and DDFT. Therefore, we begin with the definitions of these operations. Below, we follow the methods developed previously (Chechetkin & Turygin, 1994, 1995; Lobzin & Chechetkin, 2000; Chechetkin & Lobzin, 2017, 2020a, b).

### 2.1. DFT and DDFT: main definitions and relationships

The DFT harmonics corresponding to the nucleotides of type $\alpha \in (A, C, G, T)$ in a genomic sequence of length $M$ are calculated as

$$\rho_\alpha(q_n) = M^{-1/2} \sum_{m=1}^{M} \rho_{m,\alpha} e^{-i q_n m}, \quad q_n = 2\pi n / M, \quad n = 0, 1, ..., M-1. \tag{1}$$

Here $\rho_{m,\alpha}$ indicates the position occupied by the nucleotide of type $\alpha$; $\rho_{m,\alpha} = 1$ if the nucleotide of type $\alpha$ occupies the $m$-th site and 0 otherwise. The amplitudes of Fourier harmonics (or structure factors) are defined as

$$F_{\alpha\alpha}(q_n) = \rho_\alpha(q_n) \rho_\alpha^*(q_n), \tag{2}$$

where the asterisk denotes the complex conjugation. Taking into account the symmetry relationship for the structure factors, the analysis of their spectra can be restricted by the range from $n = 1$ to

$$N = [M/2],  \quad (3)$$

where the brackets denote the integer part of the quotient. The structure factors will always be normalized on the mean spectral values, which are determined by the exact sum rules,

$$f_{\alpha\alpha}(q_n) = F_{\alpha\alpha}(q_n)/\overline{F}_{\alpha\alpha}; \quad \overline{F}_{\alpha\alpha} = N_\alpha(M - N_\alpha)/M(M-1), \quad (4)$$

where $N_\alpha$ is the total number of the nucleotides of type α in a sequence of length $M$. Below, we will also use the sums,

$$S_{\alpha\beta}(q_n) = f_{\alpha\alpha}(q_n) + f_{\beta\beta}(q_n); \alpha \neq \beta, \quad (5)$$

$$S_{\alpha\beta\gamma}(q_n) = f_{\alpha\alpha}(q_n) + f_{\beta\beta}(q_n) + f_{\gamma\gamma}(q_n); \alpha \neq \beta \neq \gamma, \quad (6)$$

$$S_4(q_n) = f_{AA}(q_n) + f_{CC}(q_n) + f_{GG}(q_n) + f_{TT}(q_n), \quad (7)$$

which can be applied to the detection of quasi-periodic patterns or motifs composed of the nucleotides of different types. The period $p$ is measured in terms of the number of nucleotides (these units will always be tacitly implied below) and is calculated as,

$$p = M/n. \quad (8)$$

The harmonics in DDFT are calculated as

$$\Phi_\alpha(\tilde{q}_{n'}) = (N-1)^{-1/2} \sum_{n=2}^{N} f_{\alpha\alpha}(q_n) e^{-i\tilde{q}_{n'}n}, \quad \tilde{q}_{n'} = 2\pi n'/(N-1), \quad n' = 0, 1, ..., N-2, \quad (9)$$

where $N$ is defined by Eq. (3) and $f_{\alpha\alpha}(q_n)$ are the normalized structure factors (see Eq. (4)). The similar transform can be used for the sums defined by Eqs. (5)–(7). The amplitudes of harmonics are given by

$$F_{\alpha\alpha, II}(\tilde{q}_{n'}) = \Phi_\alpha(\tilde{q}_{n'})\Phi_\alpha^*(\tilde{q}_{n'}). \quad (10)$$

Similarly to DFT, the analysis of the spectra for the amplitudes defined by Eq. (10) can be restricted from $n' = 1$ to

$$N' = [(N-1)/2]. \quad (11)$$

The DDFT amplitudes are normalized as

$$f_{II}(\tilde{q}_{n'}) = F_{II}(\tilde{q}_{n'})/\overline{F}_{II}, \quad (12)$$

$$\overline{F}_{II} = \frac{1}{N'} \sum_{n'=1}^{N'} F_{II}(\tilde{q}_{n'}) \ . \tag{13}$$

Generally, equidistant series in DFT spectra also generate the corresponding equidistant series in DDFT spectra with the spectral numbers $k'n'$, $k' = 1, ..., k'_{max}$; $k'_{max} n' \leq N'$, where $N'$ is defined by Eq. (11). The number of quasi-periodic patterns can be assessed by the spectral number $n'$ for the peak amplitude $f_{II}(\tilde{q}_{n'})$ as

$$N'_p = (N-1)/n' \ , \tag{14}$$

while their periods in nucleotides are given by

$$p'_{II} = M/N'_p \ . \tag{15}$$

### 2.2. Correlation functions

The nucleotide correlation functions (NCF) are determined as,

$$K_{\alpha\alpha}(m_0) = M^{-1} \sum_{m=1}^{M} \rho^c_{m,\alpha} \rho^c_{m+m_0,\alpha}, \ m_0 = 0, 1, ..., M-1 \ , \tag{16}$$

$$\rho^c_{m,\alpha} = \begin{cases} \rho_{m,\alpha}, & \text{if } 1 \leq m \leq M; \\ \rho_{m-M,\alpha}, & \text{if } M+1 \leq m \leq 2M-1. \end{cases} \tag{17}$$

The circular NCFs used in this paper are especially suitable for the detection of periodic patterns. Periodic patterns with a period $p$ produce a series of equidistant peaks at the multiple spacings, $m_0 = kp$, $k = 1, 2, ...$ The corresponding mean value is given by

$$\overline{K}_{\alpha\alpha} = \frac{1}{M-1} \sum_{m_0=1}^{M-1} K_{\alpha\alpha}(m_0) = \frac{N_\alpha(N_\alpha - 1)}{M(M-1)} \ . \tag{18}$$

The correlation functions are symmetrical,

$$K_{\alpha\alpha}(m_0) = K_{\alpha\alpha}(M - m_0) \ . \tag{19}$$

This allows us to restrict the analysis of NCF from $m_0 = 1$ to $N$ defined by Eq. (3). The normalized deviations,

$$\kappa_{\alpha\alpha}(m_0) = \left( K_{\alpha\alpha}(m_0) - \overline{K}_{\alpha\alpha} \right) / <\Delta K^2_{\alpha\alpha}>^{1/2}_{random} \ , \tag{20}$$

where

$$<\Delta K^2_{\alpha\alpha}>_{random} = \overline{F}^2_{\alpha\alpha}/M, \ \overline{F}_{\alpha\alpha} = N_\alpha(M-N_\alpha)/M(M-1) \ , \tag{21}$$

are Gaussian for the random sequences. Similarly to the sums defined by Eqs. (5)–(7), it is useful to introduce the combinations,

$$Q_{\alpha\beta}(m_0) = \left(\kappa_{\alpha\alpha}(m_0) + \kappa_{\beta\beta}(m_0)\right)/\sqrt{2}; \ \alpha \neq \beta \ , \tag{22}$$

$$Q_{\alpha\beta\gamma}(m_0) = \left(\kappa_{\alpha\alpha}(m_0) + \kappa_{\beta\beta}(m_0) + \kappa_{\gamma\gamma}(m_0)\right)/\sqrt{3}; \ \alpha \neq \beta \neq \gamma \ , \tag{23}$$

$$Q_4(m_0) = \left(\kappa_{AA}(m_0) + \kappa_{TT}(m_0) + \kappa_{CC}(m_0) + \kappa_{GG}(m_0)\right)/2 \ , \tag{24}$$

which are also Gaussian for the random sequences.

The correlation functions and the DFT structure factors are not independent and are related by the Wiener-Khinchin relationship,

$$K_{\alpha\alpha}(m_0) = M^{-1} \sum_{n=0}^{M-1} F_{\alpha\alpha}(q_n) \exp(-iq_n m_0) \ . \tag{25}$$

The normalized deviations for NCF can be expressed as,

$$\kappa_{\alpha\alpha}(m_0) \equiv \Delta k_{\alpha\alpha}(m_0)/(1/M)^{1/2} \ , \tag{26}$$

$$\Delta k_{\alpha\alpha}(m_0) = (K_{\alpha\alpha}(m_0) - \overline{K}_{\alpha\alpha})/\overline{F}_{\alpha\alpha} = M^{-1} \sum_{n=1}^{M-1} \left(f_{\alpha\alpha}(q_n) - 1\right) e^{-iq_n m_0} \ . \tag{27}$$

These deviations are insensitive to the nucleotide composition and genome length but may strongly depend on the dominating underlying periodicities in genomic sequences. In the viral genomes this is the triplet periodicity $p = 3$ inherent to the protein-coding regions (for a review and further references see, e.g., Lobzin & Chechetkin, 2000; Marhon & Kremer, 2011). The relationship defined by Eq. (27) facilitates the control of contribution from underlying periodicities into the normalized deviations for NCF by cutting-off dominating peaks and re-normalizing DFT spectra. Such a procedure can be used for detection of the weaker longer periodicities on the background of strong short periodicities.

*2.3. Statistical criteria*

Throughout this paper, we will use the standard statistical criteria corresponding to the probability Pr = 0.05. For the random sequences, the statistics for the DFT and DDFT normalized harmonics defined by Eqs. (4) and (12) is Rayleighian, whereas the statistics for the normalized deviations defined by Eqs. (20) and (22)–(24) is Gaussian. To study the distribution of periodic patterns over the genome, we will use a set of non-overlapping windows of width $w$. Averaging of the DFT spectra over the windows provides the corresponding periodogram, while averaging of the normalized deviations for NCF over the windows provides the corresponding correlogram (see, e.g., Marple, Jr., 1987). Averaging over windows diminishes the effects of indels on the periodicity phasing.

*2.4. Reconstruction of motifs related to quasi-periodic patterns*

The motifs related to quasi-periodic patterns are presumably the most important for practical applications. For their reconstruction, we developed a method of transitional automorphic mapping of the genome on itself (TAMGI). The algorithm for TAMGI is as follows. Let a step length $s$ be chosen (equal to the detected period of periodic patterns in the problem concerned). Then, the pairs of nucleotides ($N_m$, $N_{m+s}$) separated by the step $s$ are mutually compared when moving one-by-one site $m$ along the genomic sequence. If both nucleotides belong to the same type, they both are retained in the genomic sequence; otherwise, the nucleotide $N_m$ is replaced by void (denoted traditionally by the hyphen). Thus, the $N_m$-th nucleotide will be retained if it has at least one neighbor $N_{m-s}$ or $N_{m+s}$ of the same type and be replaced by void otherwise. The resulting sequence after TAMGI is composed of the nucleotides of four types (A, C, G, T) and the hyphens "-" denoting voids. Further analysis is reduced to the enumeration of all complete words of length $k$ ($k$-mers) composed only of nucleotides (voids within the complete words are prohibited) and surrounded by the voids "-" at 5'- and 3'-ends, -$N_k$-. By definition, the complete words are non-overlapping. At the next stage, the mismatches to the complete words can be studied. If the presence of periodic patterns is ensured, e.g., by DFT or DDFT, TAMGI with the step $s$ equal to the corresponding period $p$ provides a sequence enriched by the periodic patterns. Thus, TAMGI contains the most frequent motifs related to quasi-periodic patterns and provides their distribution over the genome. As TAMGI contains also the quasi-random fraction, the latter can be partially filtered out by combining TAMGI with the steps $s$ and $2s$. The TAMGI method is robust with respect to indels but may depend on the nucleotide content and underlying short periodicities.

Generally, TAMGI may also be extended to non-integer steps $s$ by the best integer approximation of transitional mapping with non-integer $s$. The latter can be obtained using a set of chains ($N_1$, $N_{1+\{s\}}$, ..., $N_{1+\{k_{max}s\}}$), ($N_2$, $N_{2+\{s\}}$, ..., $N_{2+\{k_{max}s\}}$), ($N_{\{s\}}$, $N_{\{2s\}}$, ..., $N_{\{(k_{max}+1)s\}}$), where $\{ks\}$ means rounding to the nearest integer and $\{(k_{max}+1)s\} < M$. The choice of consecutive pairs in the chains is performed by the algorithm similar to that as described above.

*2.5. Spectral entropy*

The abundance of the genomic DNA/RNA sequences by quasi-periodic patterns can be assessed by the spectral entropy (Chechetkin & Turygin, 1994; Chechetkin & Lobzin, 1996; Chechetkin, 2011; Balakirev et al., 2003, 2005, 2014). The spectral entropy is defined as,

$$S_\alpha = -\sum_{n=1}^{N} f_{\alpha\alpha}(q_n)\ln f_{\alpha\alpha}(q_n); \quad S_{total} = \sum_\alpha S_\alpha . \tag{28}$$

Its mean value,

$$<S_\alpha>_{random} = -(1-C)N , \tag{29}$$

where $C$ is Euler constant; $(1-C) = 0.422785...$, attains approximate maximum for the random sequences. The corresponding variance for the spectral entropy is given by

$$\sigma^2(S_\alpha)_{random} = 0.289868...N . \tag{30}$$

The abundance of quasi-periodic patterns in the genomes of different lengths can be assessed by the relative spectral entropies,

$$S_{\alpha, rel} = S_\alpha / |<S_\alpha>_{random}| . \tag{31}$$

The relative spectral entropy serves also for the assessment of the load from point mutations and indels on the genomes or on the particular genes and pseudogenes (Balakirev et al., 2003, 2005, 2014).

## 3. Results

*3.1. Nucleocapsid structure and packaging of genomic ssRNA*

Early studies based on electron microscopy have revealed that the ribonucleocapsid of coronaviruses is helical, consisting of coils of 9–16 nm in diameter and a hollow interior of approximately 3–4 nm (Macneughton et al., 1978). Chang et al. (2014) asserted that for the SARS-CoV nucleocapsid an outer diameter of 16 nm and an inner diameter of 4 nm are consistent with cryo-EM observations. The length of a helical turn per pitch is

$$l_t = \left(1 + (\pi d/h)^2\right)^{1/2} h , \tag{32}$$

where $d$ is the diameter of the helix and $h$ is the pitch. According to Chen et al. (2007), the pitch for the SARS-CoV nucleocapsid is $h = 14$ nm. Taking the distance between RNA bases as 0.34 nm, the positioning of RNA near the inner diameter of nucleocapsid provides the length of RNA turn about 54–56 nt, the positioning of RNA in the middle between the inner and outer diameters would provide the length of turn about 84–87 nt, whereas the positioning of RNA at the outer diameter would provide the length of turn about 153–154 nt. Chang et al. (2009) found multiple (at least three) nucleic acid binding sites in N proteins. Therefore, the intermediate dynamic positioning of RNA in the middle during assembly/packaging cannot be excluded. At the final stage of packaging, ssRNA is assumed to be positioned at the inner diameter of the nucleocapsid in accordance with cryo-EM observations (Chang et al., 2014). We performed the complete combined analysis of the SARS-CoV and SARS-CoV-2 genomes based on DFT, DDFT, NCF, and pattern correlation functions (Chechetkin & Lobzin, 2020b) and screened all range of putative periods from the shortest period of 2 nt to the large-scale periods comparable to the whole genome lengths. The most interesting results related to the ribonucleocapsid assembly/packaging are presented below.

### *3.2. Overview of NCF and characteristic patterns*

We took for analysis one genomic sequence for SARS-CoV (GenBank accession: NC_004718; $M = 29751$, $N_A = 8481$, $N_G = 6187$, $N_T = 9143$, $N_C = 5940$) as a reference and the genomic sequences for three isolates of SARS-CoV-2 (GenBank accessions: MT371038; $M = 29719$, $N_A = 8873$, $N_G = 5834$, $N_T = 9554$, $N_C = 5458$; MT295464; $M = 29892$, $N_A = 8948$, $N_G = 5862$, $N_T = 9592$, $N_C = 5490$; MT371037; $M = 29694$, $N_A = 8866$, $N_G = 5829$, $N_T = 9544$, $N_C = 5455$) to assess the impact of point mutations and indels on the detected patterns. Henceforth, the viruses will be denoted by their accessions. Taking into account the transitional invariance of a helix, the main results will be given for NCF. The presence of periodic components in NCF was proved by combining DFT and DDFT. The general overviews of the plots for the normalized NCF deviations defined by Eq. (20) are shown in Figs. 1 and 2. The overview for MT371038 is closer to that shown in Fig. 1, while the corresponding plots for MT295464 are similar to those shown in Fig. 2.

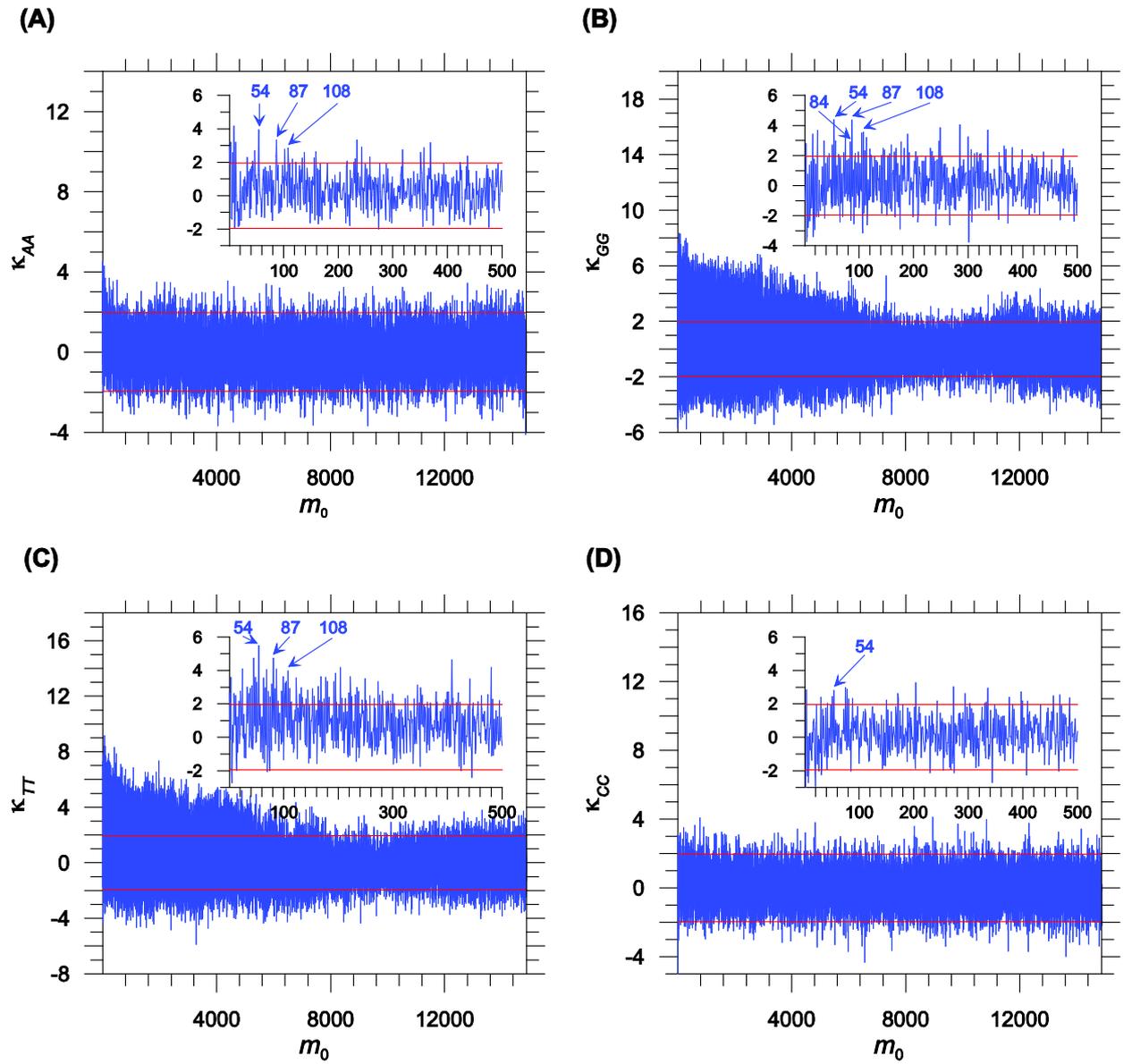

**Figure 1.** The plots for the normalized NCF deviations defined by Eqs. (26) and (27). The initial ranges of plots shown in the inserts were re-calculated by replacing the highest Fourier harmonics by the peaks defined by extreme value statistics in the DFT spectra. The characteristic spacings $m_0$ are explicitly marked by the arrows. The horizontal lines correspond to the significance Pr = 0.05 for the reshuffled random sequences. The panels A–D correspond to the nucleotides of particular types in the genome of SARS-CoV (accession NC_004718).

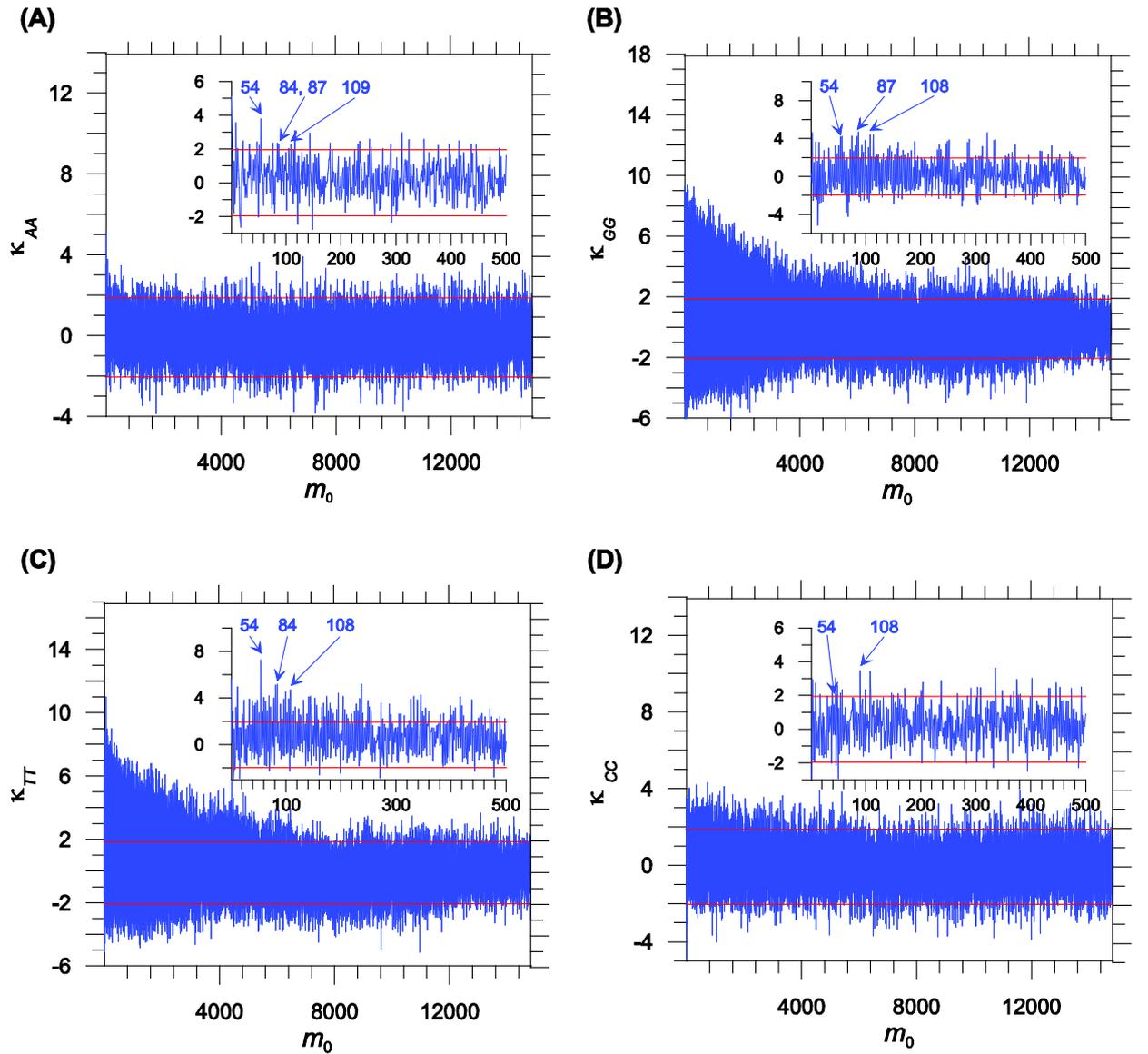

**Figure 2.** The plots for the normalized NCF deviations defined by Eqs. (26) and (27). The initial ranges of plots shown in the inserts were re-calculated by replacing the highest Fourier harmonics by the peaks defined by extreme value statistics in the DFT spectra. The characteristic spacings $m_0$ are explicitly marked by the arrows. The horizontal lines correspond to the significance Pr = 0.05 for the reshuffled random sequences. The panels A–D correspond to the nucleotides of particular types in the genome of SARS-CoV-2 (accession MT371037).

Then, all plots for NCF were recalculated using Eqs. (26) and (27) and replacing all highest harmonics in the DFT spectra by the peaks assessed by extreme value statistics (cf. Chechetkin & Lobzin, 2019). The initial ranges of the recalculated plots are shown in the inserts to Figs. 1 and 2. The deviations corresponding to the most pronounced patterns are shown explicitly by arrows. Such patterns are quasi-periodic because the corresponding approximately equidistant series can be pursued in these plots (the next peaks are shown only for the most pronounced patterns with periodicity $p = 54$).

*3.3. Correlograms and periodograms for the genomes of SARS-CoV and SARS-CoV-2*

For the further analysis and as a cross-check of the above results, the NCF and DFT spectra were computed for the set of non-overlapping windows of width 432 nt. The 3'-end windows #69 were incomplete for the genomes of NC_004718, MT371038, and MT371037. The characteristics used in our analysis are robust with respect to the length of window. The normalized deviations for NCF were calculated using Eqs. (26) and (27) and replacing peaks corresponding to the triplet periodicity $p = 3$ by the heights corresponding to Pr = 0.05 in the Rayleigh spectra. The similar cut-off was used after the calculations of the DFT spectra within windows. The correlograms obtained by the averaging of the plots for normalized NCF deviations for the sums defined by Eq. (24) are shown in Fig. 3. The significance threshold of Pr = 0.05 for the correlograms corresponds to $\pm 1.96/N_w^{1/2}$, where $N_w$ is the total number of windows. In all genomes the deviations for $m_0 = 54$ were the highest and the deviations with $m_0 = 108$ were significant as well. For SARS-CoV the deviations with $m_0 = 216$ (=4×54) were also significant. The next characteristic high deviations for SARS-CoV were for $m_0 = 87$, while in the isolates of SARS-CoV-2 they were for $m_0 = 84$.

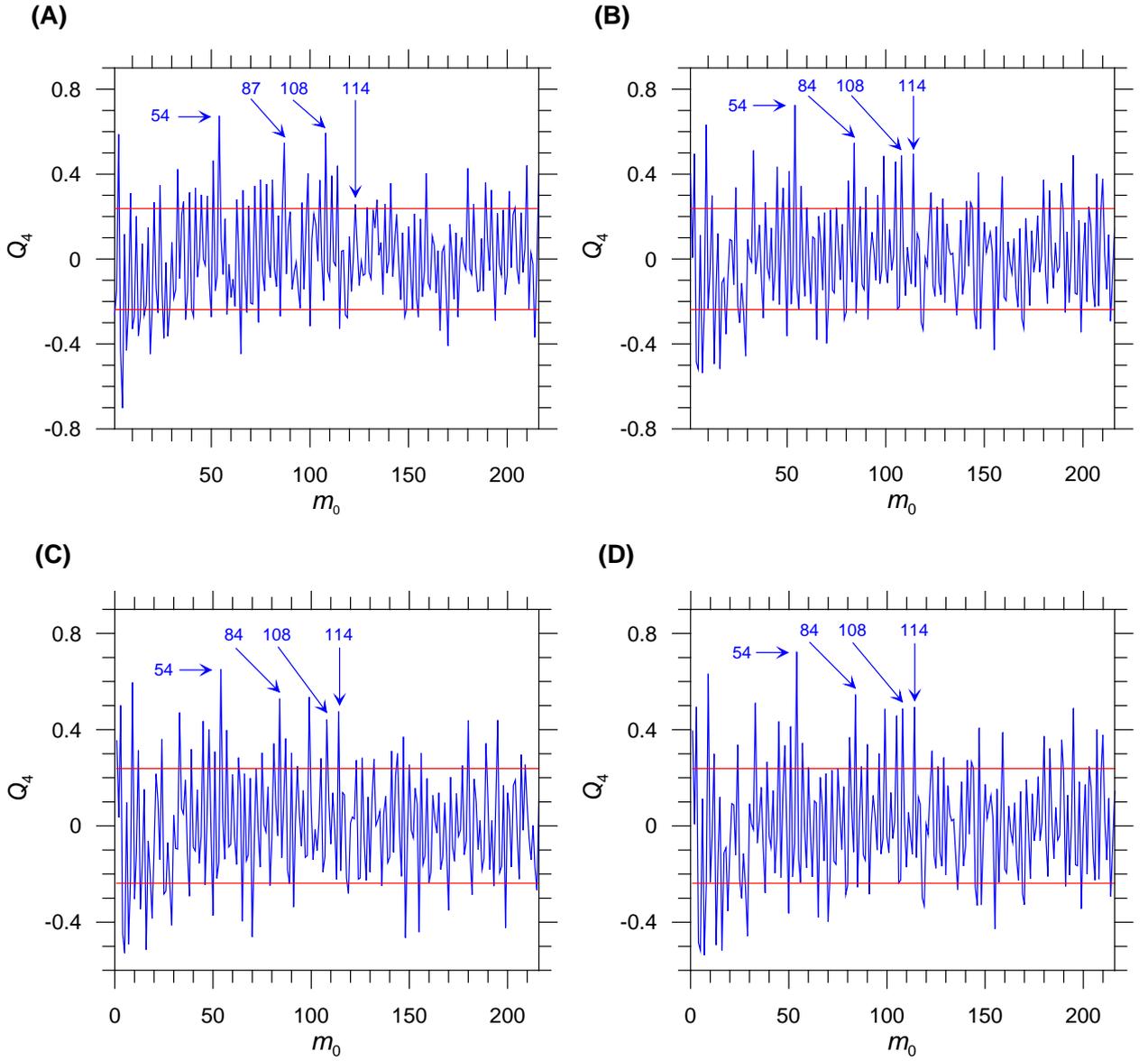

**Figure 3.** The correlograms obtained by the averaging of the normalized NCF deviations defined by Eqs. (24), (26), and (27) calculated within non-overlapping windows of width 432 nt for the genomes of NC_004718 (A), MT371038 (B), MT295464 (C), and MT371037 (D). The horizontal lines correspond to the significance Pr = 0.05.

The corresponding periodograms obtained by the averaging of the DFT spectra over windows were then re-computed by DDFT (Chechetkin & Lobzin, 2020a). Due to the restrictions related to the applicability of DDFT, the left boundary in the DDFT spectra is positioned at $n' = 10$. Then, the DDFT spectra were renormalized in this range. The resulting DDFT spectra for the sums defined by Eq. (7) are shown in Fig. 4. Again, the harmonic with $n' = 27$, $p' = 54.2$ was reproducibly significant and the highest in the range under study for all genomes. For SARS-CoV the harmonic with $n' = 43$, $p' = 86.4$ was also significant, whereas for the isolates of SARS-CoV-2 the harmonic with $n' = 42$, $p' = 84.4$ appeared to be insignificant. The harmonic with $n' = 57$, $p' = 114.5$ for SARS-CoV can be treated as a distorted and modified doubled period $p = 54$ (typically of hidden fuzzy repeating patterns). Thus, combining correlograms for NCF with the analysis of periodograms by DDFT reveals clearly the persistently reproducible quasi-periodic patterns with the period $p \approx 54$ in all genomes and indicates the relevance of less robust patterns with $p \approx 84$ and $87$.

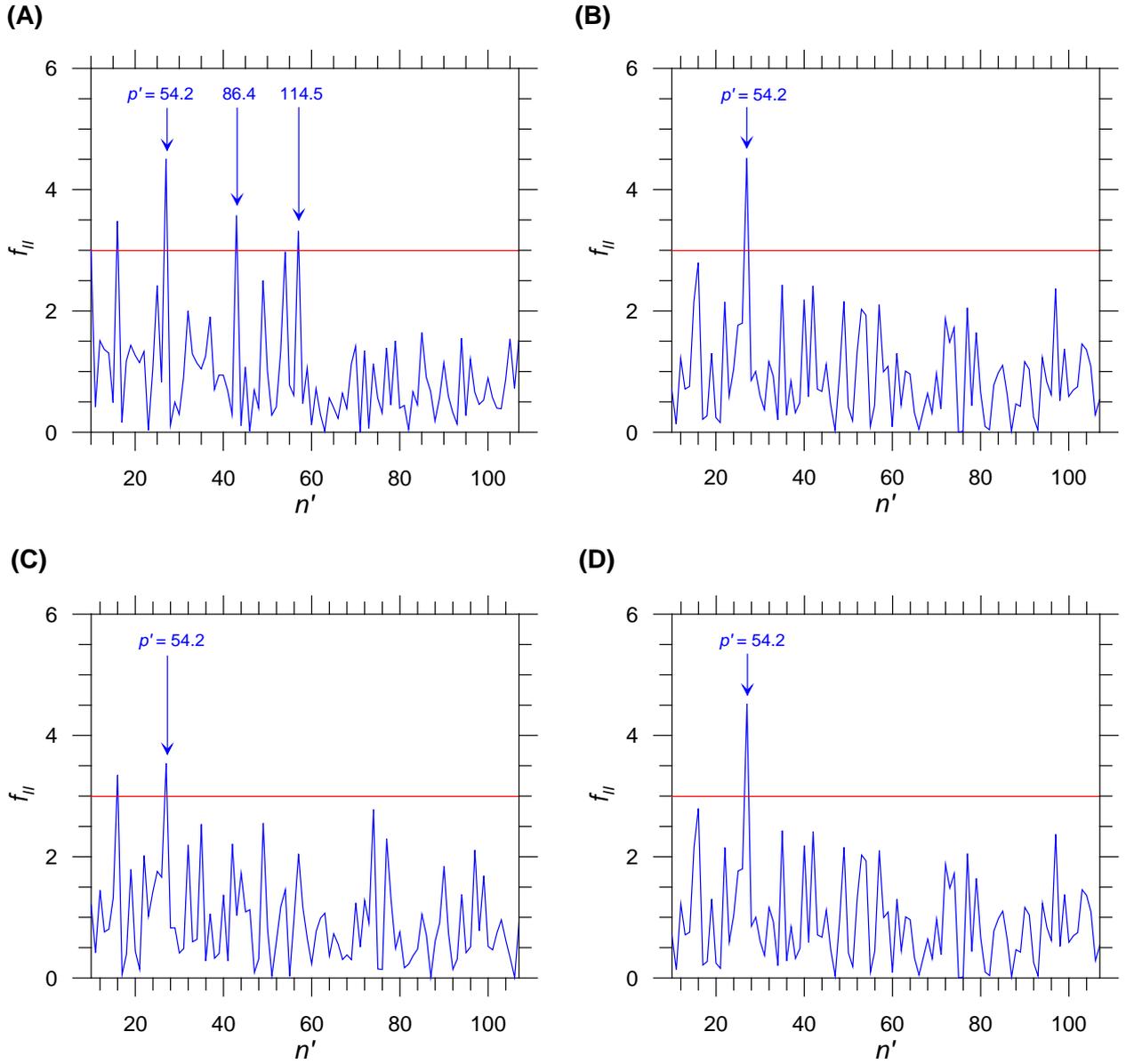

**Figure 4.** The DDFT spectra of the periodograms obtained by averaging of the DFT spectra defined by Eqs. (4) and (7) calculated within non-overlapping windows of width 432 nt for the genomes of NC_004718 (A), MT371038 (B), MT295464 (C), and MT371037 (D). The horizontal lines correspond to the significance $Pr = 0.05$.

### *3.4. Distribution over the genomes for deviations of NCF components putatively related to ribonucleocapsid assembly/packaging signals*

To assess the distribution of the detected patterns over the genomes, the normalized deviations for NCF were computed in separate windows of width 432 nt as described above. The spacings for NCF $m_0$ were chosen by the correspondence with the periods of the detected patterns and were equal to 54, 84, and 87, respectively. The resulting plots for the sums defined by Eq. (24) are shown in Fig. 5. The numerical data for the profiles in Fig. 5 and for the profiles corresponding to the nucleotides of particular types as well as to the sums defined by Eqs. (22) and (23) are collected in Supplement S1. We assessed the correlations between different profiles by the Pearson correlation coefficients. The NCF profiles for the different genomes were significantly correlated for the same spacings $m_0$, while the profiles with the different spacings can be considered uncorrelated. The coefficients for correlations between profiles for SARS-CoV and three isolates of SARS-CoV-2 at $m_0 = 54$ were 0.623, 0.491, and 0.636 (Pr $< 2 \times 10^{-5}$ for 69 components). The related coefficients for the correlations MT371038–MT295464, MT371038–MT371037, and MT295464–MT371037 were 0.817, 0.954, and 0.751. Similar but a bit lower values were obtained for the correlations at two other spacings.

As supposed, the motifs detected at the different spacings $m_0$ are related to the different stages of ribonucleocapsid assembly/packaging. Such motifs can be incorporated into the genomic sequence by silent mutations due to the degeneracy of the genetic code. The regular near-by positioning of different assembly/packaging motifs would be too restrictive, because the main function of the genomic sequence is coding for proteins. Therefore, the windows enriched simultaneously by the motifs of different types are especially interesting as well as the windows enriched or depleted by the motifs of the same type. Despite evolutionary divergence between the two viruses and the action of point mutations and indels, some features appear to be remarkably reproducible in all genomes. In particular, in the window #3 (sites 865–1296) the normalized NCF deviations exceeded significance threshold Pr = 0.05 for all genomes at $m_0 = 54$. Similar but stronger effects were observed for the window #5 (1729–2160); in the latter case for SARS-CoV, this window was also enriched by the motifs with $m_0 = 87$. The window #29 (12097–12528) was enriched by the motifs with $m_0 = 54$; additionally, for all isolates of SARS-CoV-2 this window was enriched by the motifs with $m_0 = 87$. An opposite example with depletion of motifs associated with $m_0 = 87$ can be seen in the window #34 (14257–14688). These profiles may explain why the mean deviation with $m_0 = 84$ exceeds the deviation with $m_0 = 87$ in the genomes of SARS-CoV-2. In the latter case, despite significant enrichment by the motif with $m_0 = 87$ in some of windows, there are also the windows with significant depletion of this motif.

**(A)**

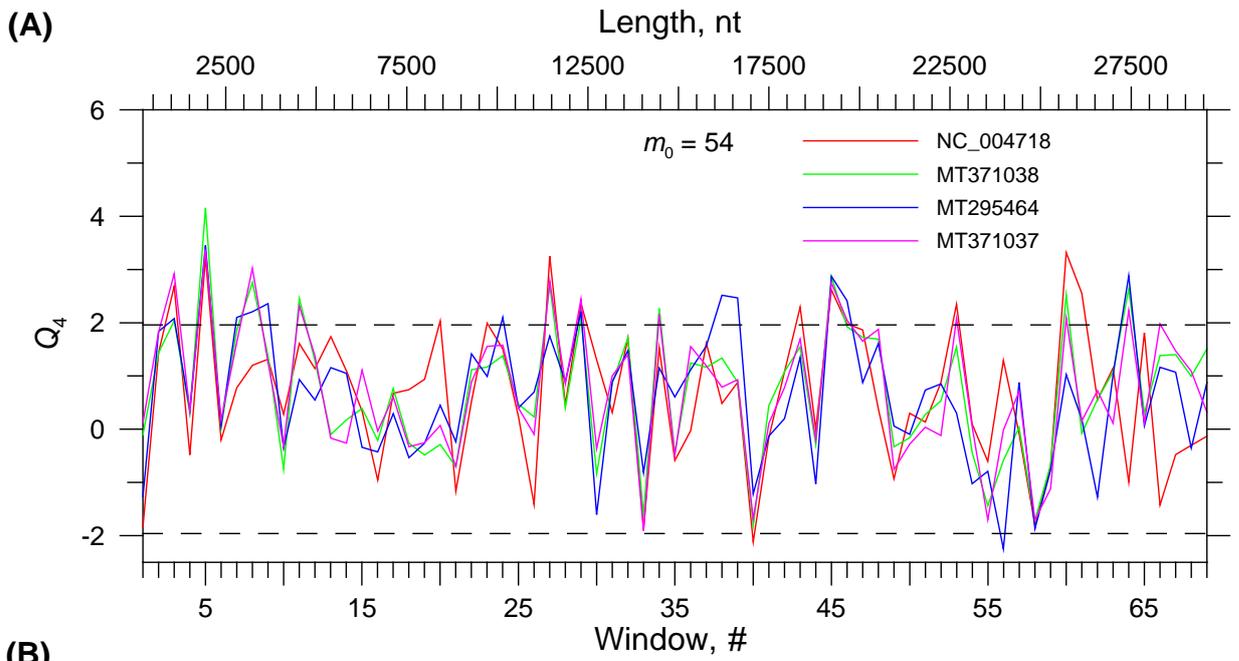

**(B)**

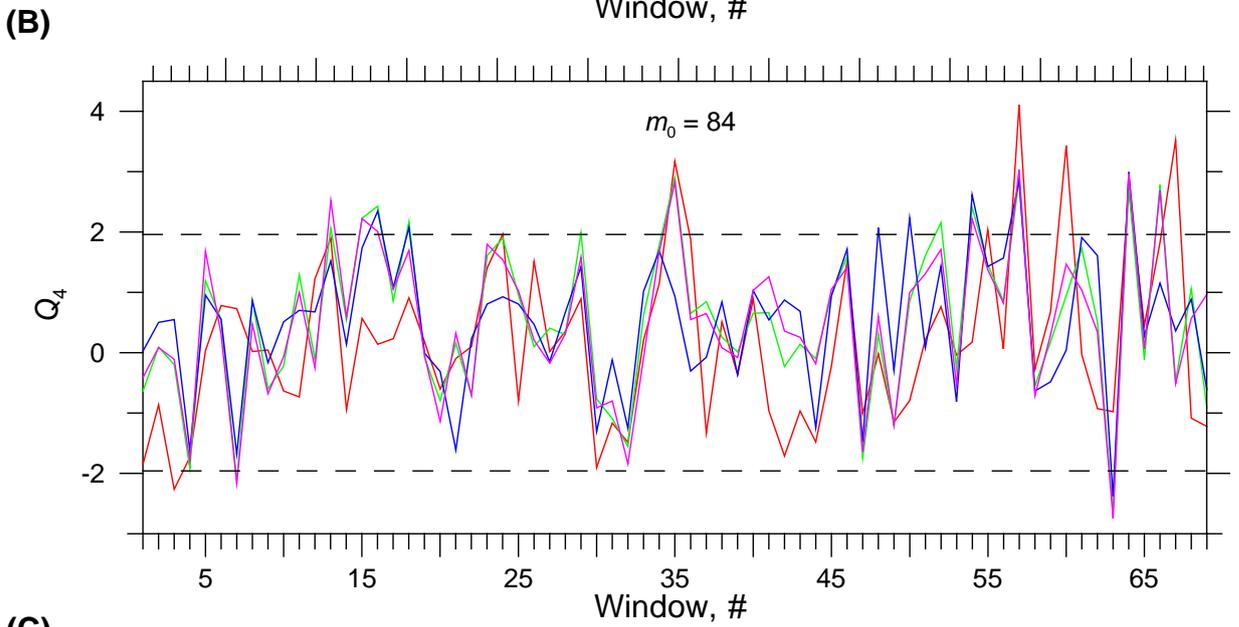

**(C)**

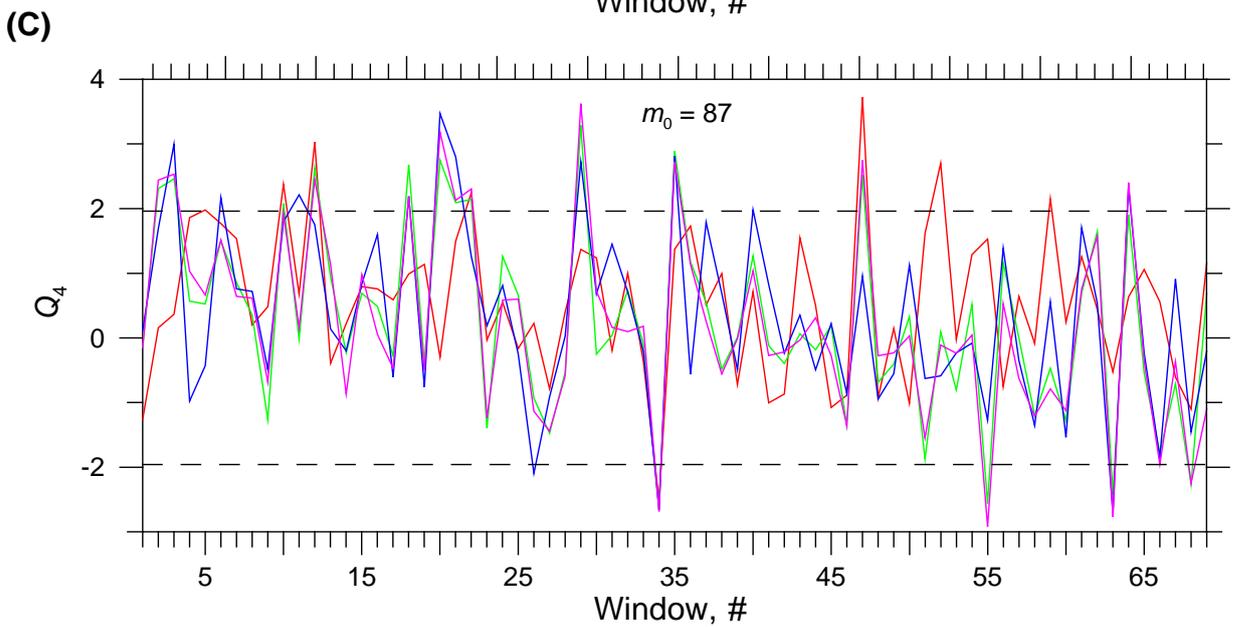

**Figure 5.** The profiles of the normalized NCF deviations defined by Eqs. (24), (26), and (27) for the components with $m_0$ = 54 (A), 84 (B), and 87 (C) calculated within non-overlapping windows of width 432 nt for the genomes of NC_004718, MT371038, MT295464, and MT371037. The horizontal lines correspond to the significance Pr = 0.05.

Similar profiles were also obtained for DDFT harmonics with the spectral numbers $n'$ = 27, 42, and 43. DDFT spectra in windows of width 432 were computed for the sum defined by Eq. (7) as described above. The related profiles can be found in Supplement S2. The counterpart profiles for the normalized NCF deviations and DDFT harmonics appear to be significantly correlated in the same genomes. Therefore, the characteristic features in the both sets of profiles were approximately reproducible. In addition to these features, an extremely high peak for the DDFT harmonic with $n'$ = 27, $p'$ = 54.2 was observed in the window #60 (25489–25920) in the genome of SARS-CoV.

### 3.5. Distributions and repertoires of motifs obtained by TAMGI

Reconstructed motifs and their positions on the genomes were obtained by TAMGI with the steps $s$ = 54, 84, and 87. The resulting sequences after TAMGI are explicitly reproduced in Supplements S3–S6. The data on the total fractions of nucleotides after TAMGI are summarized in Table 1. A simple theoretical consideration shows that the partial fractions of nucleotides after TAMGI for the randomly reshuffled genomic sequences are given by

$$\Phi_\alpha = \varphi_\alpha^2 (2 - \varphi_\alpha); \; \Phi_{total} = \sum_\alpha \Phi_\alpha , \tag{33}$$

where $\varphi_\alpha$ is the frequency of nucleotides of the type α retained under reshuffling. Eq. (33) was additionally verified by simulations. The frequencies given by Eq. (33) are independent of steps and also are reproduced in Table 1 for reference. The variances of frequencies related to particular random realizations are about

$$\sigma^2(\Phi_\alpha) = \Phi_\alpha (1 - \Phi_\alpha)/M; \; \sigma^2_{total} = \sum_\alpha \sigma^2(\Phi_\alpha) . \tag{34}$$

Eq. (34) yields for $\sigma_{total}$ the value of 0.004 that is much lower than the differences between frequencies for viral and random sequences. In this sense, Table 1 reveals distinctly non-random character of the variations related to the detected quasi-periodic patterns in the viral genomes. The mutual comparison of the total frequencies of nucleotides after TAMGI for the different isolates of SARS-CoV-2 shows their robustness against point mutations and indels.

**Table 1**

The total frequencies of nucleotides after TAMGI with different steps in the genomes of SARS-CoV and SARS-CoV-2

| | Genome accession | | | |
|---|---|---|---|---|
| | NC_004718 | MT371038 | MT295464 | MT371037 |
| Random | 0.448 | 0.456 | 0.456 | 0.456 |
| $s=54$ | 0.481 | 0.490 | 0.489 | 0.490 |
| $s=84$ | 0.464 | 0.483 | 0.483 | 0.483 |
| $s=87$ | 0.476 | 0.477 | 0.477 | 0.477 |

The total frequencies of nucleotides after TAMGI for the randomly reshuffled genomic sequences were calculated by Eq. (33).

The general distributions of $k$-mers, -$N_k$-, on the length $k$ are presented in Table 2. The period of $p \approx 54$ implies the association of 6.75 nt per one N protein (see Section 4.2 below). All motifs with $k \geq 6$ and their positions on the genomes are enumerated in Supplement S7. The profiles for the total numbers of nucleotides within non-overlapping windows of width 432 nt after TAMGI with the steps $s = 54$, 84, and 87 are shown in Fig. 6. For the incomplete windows #69 these numbers were increased proportionally to obtain estimates for the width of 432 nt. The profiles in Figs. 5 and 6 are close but differ in some features. The corresponding Pearson correlation coefficients between the counterpart profiles in Figs. 5 and 6 were highly significant, 0.72–0.86. Nevertheless, the highest peaks and the lowest valleys interesting from the point of view of applications may differ between the counterpart profiles. In particular, the highest peak in Fig. 6A was observed for the window #8 (sites 3025–3456). The profiles for $s = 54$ and 87 were slightly biased from the higher values at 5'-end to the lower values at 3'-end, though the extreme windows #1 (1–432) comprising 5'-UTR were depleted of motifs. The numerical values for all profiles in Fig. 6 can be found in Supplement S7.

**Table 2**

The occurrences of $k$-mers, -$N_k$-, in the genomes of SARS-CoV and SARS-CoV-2 after TAMGI with steps $s$=54, 84, and 87

| $k$-mers | Genome accession | | | |
|---|---|---|---|---|
| | NC_004718 | MT371038 | MT295464 | MT371037 |
| | | $s=54$ | | |
| 1 | 3618 | 3617 | 3633 | 3609 |
| 2 | 1834 | 1823 | 1847 | 1828 |
| 3 | 862 | 898 | 901 | 901 |
| 4 | 458 | 448 | 449 | 446 |
| 5 | 218 | 226 | 227 | 226 |
| 6 | 106 | 128 | 125 | 126 |
| 7 | 51 | 51 | 52 | 52 |
| 8 | 28 | 23 | 23 | 23 |
| 9 | 16 | 21 | 20 | 20 |
| 10 | 5 | 10 | 10 | 10 |
| 11 | 4 | 2 | 2 | 2 |
| 12 | 5 | 2 | 2 | 2 |
| 13 | 1 | – | – | – |
| 14 | – | 1 | 1 | 1 |
| 17 | – | 1 | 1 | 1 |
| | | $s=84$ | | |
| 1 | 3844 | 3790 | 3808 | 3787 |
| 2 | 1782 | 1807 | 1819 | 1802 |
| 3 | 870 | 899 | 903 | 901 |
| 4 | 368 | 417 | 422 | 416 |
| 5 | 211 | 197 | 198 | 197 |
| 6 | 93 | 108 | 108 | 108 |
| 7 | 49 | 50 | 51 | 51 |
| 8 | 20 | 35 | 35 | 35 |
| 9 | 12 | 13 | 12 | 12 |
| 10 | 4 | 10 | 10 | 10 |
| 11 | 2 | 7 | 7 | 7 |
| 12 | 1 | 1 | 1 | 1 |
| 13 | – | 1 | 1 | 1 |
| 15 | 2 | – | – | – |
| | | $s=87$ | | |
| 1 | 3667 | 3726 | 3752 | 3732 |
| 2 | 1855 | 1798 | 1816 | 1793 |
| 3 | 941 | 905 | 908 | 902 |
| 4 | 405 | 417 | 420 | 415 |
| 5 | 197 | 196 | 194 | 195 |
| 6 | 102 | 113 | 116 | 115 |
| 7 | 52 | 44 | 44 | 44 |
| 8 | 23 | 33 | 31 | 32 |
| 9 | 8 | 9 | 10 | 9 |
| 10 | 6 | 9 | 9 | 9 |
| 11 | 3 | 5 | 5 | 5 |
| 12 | 2 | 1 | 1 | 1 |
| 13 | – | 1 | 1 | 1 |

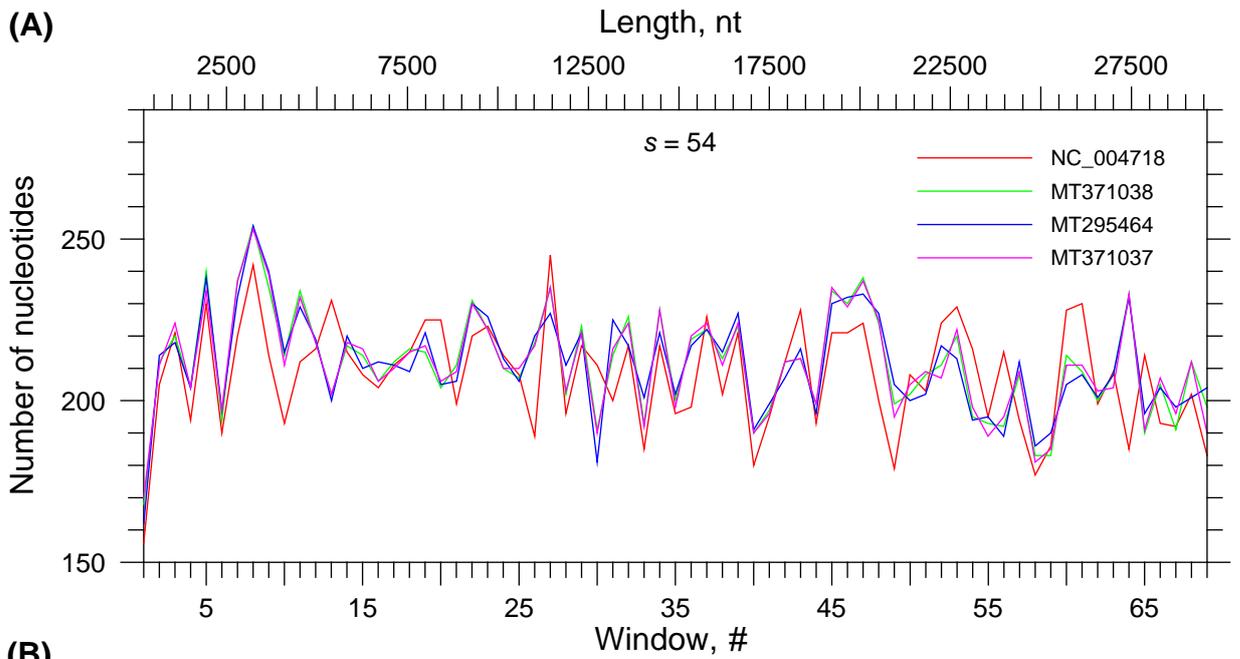
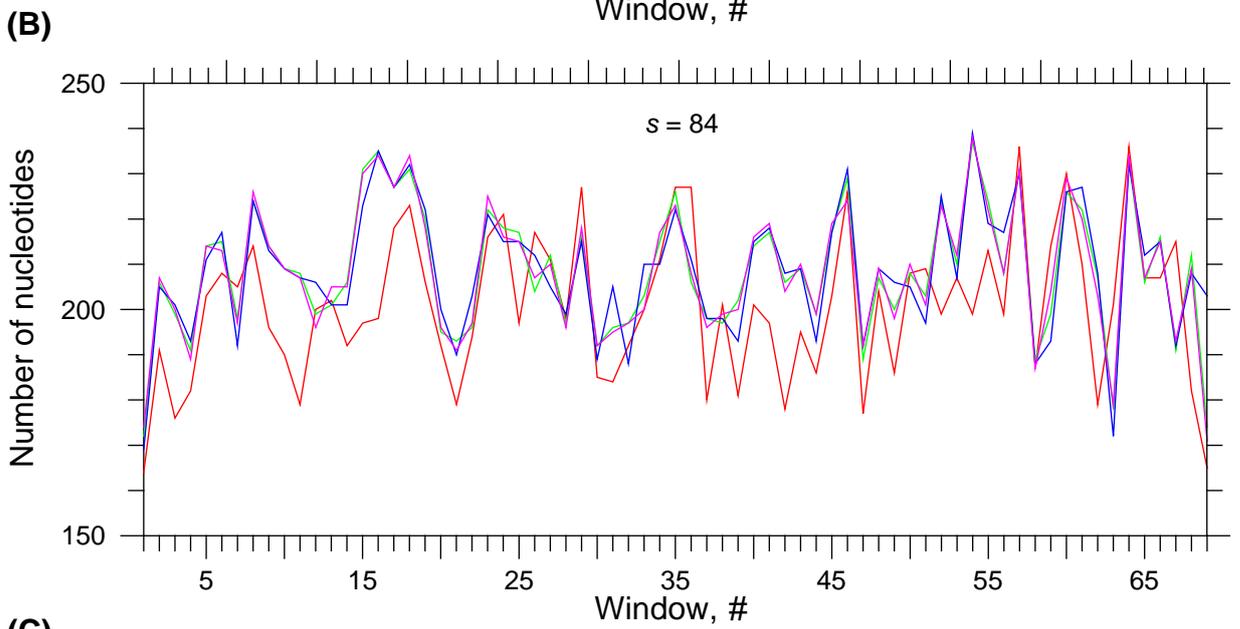
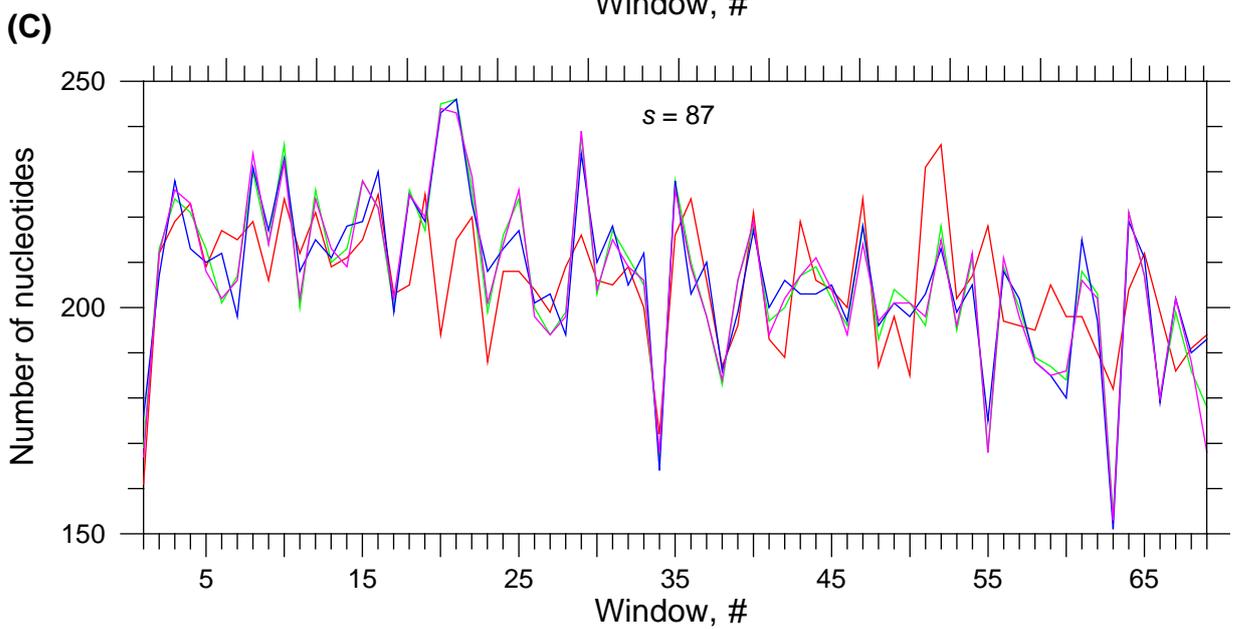

**Figure 6.** The profiles of the total numbers of nucleotides after TAMGI with steps *s* = 54 (A), 84 (B), and 87 (C) within non-overlapping windows of width 432 nt for the genomes of NC_004718, MT371038 , MT295464, and MT371037.

The comparison of repertoires of motifs with $k \geq 6$ presented in Supplement S7 revealed nearly complete correspondence (up to one-two motifs) between motifs for three isolates of SARS-CoV-2. The divergence between motifs for SARS-CoV and SARS-CoV-2 appeared to be more significant. In particular at the step *s* = 54, only 22 hexamer motifs from 102 different motifs (106 in total) in the SARS-CoV genome coincided with those for SARS-CoV-2 and 36 hexamers differed by one letter from the repertoires of hexamers for SARS-CoV-2. The similar comparison for the other steps yielded the coincidence of 18 from 89 different motifs (93 in total) and 38 motifs differing by one letter at *s* = 84 and the coincidence of 15 from 93 different motifs (102 in total) and 37 motifs differing by one letter at *s* = 87. This means that the repertoires of relatively long motifs are robust to point mutations and indels for the separate coronaviruses but diverge (and in this sense are specific enough) between the two viruses despite the conservation of the main helical periodicity $p \approx 54$ nt. The relationships between motifs found for the assembly/packaging and the other *cis*-acting elements (Madhugiri et al., 2016) should be established separately. Our study showed that actually any *cis*-acting element should comprise contextual surrounding vicinity of several tens of nucleotides up- and downstream the element.

The occurrences of the motifs determined by TAMGI can be compared with their counterparts in the whole genome. The statistical significance of such motifs in the whole genome can be assessed by the related occurrences in the sequences obtained by the random reshuffling of the genome. Instead of modeling with genome reshuffling, the rigorous theory by Zubkov & Mikhailov (1974) and Karlin & Altschul (1990) can be used for the assessment of motif occurrences (see also Boeva et al., 2006; Suvorova et al., 2014).

## 4. Discussion

### *4.1. Comparison of abundance of quasi-periodic patterns in the SARS-CoV and SARS-CoV-2 genomes*

Short tandem repeats in human genomes are widely used in the medical diagnostics and forensic (see, e.g., Grover & Sharma, 2017; Baine & Hui, 2019; Sznajder & Swanson, 2019; Butler, 2011; Kayser, 2017; and references therein). Similar patterns were also found in some prokaryotic genomes (Subirana & Messeguer, 2019). Quasi-repeating patterns in viral genomes are present commonly in the hidden form on the background of frequent random point mutations and indels. Nevertheless, many quasi-repeating patterns remain persistent, robust and contain important information about molecular mechanisms of virus life cycle, including genome packaging. Such patterns can be detected and quantified by DFT, DDFT, NCF, and other methods. Surprisingly, the quasi-repeating patterns in viral genomes are usually completely ignored when discussing evolutionary and subtyping problems in virology (see, e.g., Forster et al., 2020; Cagliani et al., 2020; Tang et al., 2020; Andersen et al., 2020; MacLean et al., 2020).

**Table 3**

The relative spectral entropies (see Eq. (31)) characterizing the abundance of quasi-periodic patterns in the viral genomes

| | Relative spectral entropy | | | | |
|---|---|---|---|---|---|
| Accession | $S_{A, rel}$ | $S_{G, rel}$ | $S_{T, rel}$ | $S_{C, rel}$ | $S_{total, rel}$ |
| NC_004718 | -1.017 | -1.233 | -1.200 | -1.008 | -4.459 |
| MT371038 | -1.001 | -1.267 | -1.220 | -1.029 | -4.517 |
| MT295464 | -1.008 | -1.251 | -1.231 | -1.046 | -4.536 |
| MT371037 | -1.007 | -1.271 | -1.222 | -1.042 | -4.542 |

The standard deviations for the relative spectral entropies $S_{\alpha, rel}$ in the random sequences of the same lengths are about 0.010. The standard deviation for $S_{total, rel}$ is twice of this value.

The abundance of quasi-periodic patterns in viral genomes can be conveniently assessed by the relative spectral entropy (Section 2.5). The more negative the spectral entropy, the higher the abundance of quasi-periodic patterns in the genome. The relevant data for the genomes of SARS-CoV and three isolates of SARS-CoV-2 are summarized in Table 3. For the significance of Pr = 0.05, the difference between the total spectral entropies should exceed by the absolute value the threshold $1.96\sqrt{2}\sigma(S_{total, rel}) \approx 0.055$. This is actually fulfilled for all three differences between $S_{total, rel}$ for SARS-CoV and the isolates of SARS-CoV-2, whereas the mutual differences between $S_{total, rel}$ for the isolates of SARS-CoV-2 are less, that is in accordance with the evolutionary divergence of SARS-CoV and SARS-CoV-2. The values of $S_{total, rel}$ in Table 3 reveal the higher enrichment of the SARS-CoV-2 genomes by periodic patterns in comparison with the SARS-CoV genome. It can also be said that during virus age the load from point mutations and indels on the genome of SARS-CoV was higher in comparison with the load on the genome of SARS-CoV-2. Within such interpretation SARS-CoV-2 may be treated as a "newborn" virus.

### *4.2. How many N proteins are needed for the complete packaging of the SARS-CoV and SARS-CoV-2 ssRNA genomes?*

The periods of ssRNA turns packaged within helical ribonucleocapsid and detected via repeating motifs in the genomic RNA sequences proved to be persistent in the genomes of SARS-CoV and SARS-CoV-2, though the repertoires of related motifs appeared to be divergent. Taking into account that the turn of nucleocapsid is composed of two octamers (Chen et al., 2007) polymerized from dimeric N proteins, the detected period of 54 nt implies that one N protein should be associated with 6.75 nt. This is very close to the estimate obtained by Chang et al. (2014) that one N protein should be associated with 7 nt. Consequently, for genomes of length 30,000 nt typical of coronaviruses, $4.4 \times 10^3$ N proteins are needed for complete packaging of the genomic ssRNA. The latter estimate significantly exceeds the value suggested by Neuman & Buchmeier (2016), $0.7–2.2 \times 10^3$ N proteins per virion and the association of each N protein with 14–40 nt of genomic RNA. The flower-like packaging of the helical nucleocapsid within the envelope (see, e.g., Gui et al., 2017; Masters, 2019; and further references therein) implies an integrity of the nucleocapsid and gives evidence against rods-on-a-string model for the nucleocapsid. Therefore, such difference in estimates cannot be attributed to uncovering of a part of the genome. Presumably, the total number of N proteins per virion is underestimated and the number $4.4 \times 10^3$ makes N proteins the most abundant in the active phase of the virus life cycle.

*4.3. Implications for therapeutic targeting*

N proteins of the coronaviruses provide the promising therapeutic targets (Chang et al., 2014; 2016; Tilocca et al., 2020; Lin et al., 2020; Yadav et al., 2020). The advantages of using N proteins for therapeutic targeting are as follows. (i) As N proteins are abundant, the antibodies against them can be used for early diagnostics and in vaccines. (ii) N proteins are multifunctional and participate not only in the assembly/packaging of the ribonucleocapsid but also in the regulation of the replication-transcription processes (Hurst et al., 2010; Verheije et al., 2010; McBride et al., 2014). The interaction between M and N proteins plays an important role in the packaging of ribonucleocapsid within envelope (Kuo et al., 2016). (iii) Coronavirus M and N proteins stand out as being the most conserved among structural proteins (Neuman & Buchmeier, 2016). They should be more stable against the load from point mutations and indels especially frequent in viruses. The most of vaccines are currently developed against spike (S) proteins. However, S proteins are rather variable and in any case the multi-targeted vaccines will be more efficient in comparison with one-targeted.

The other strategy is related to the development of RNA vaccines (Kramps & Elbers, 2017) or to targeting of specific motifs in the viral RNA. The latter can be performed by RNA aptamers, RNA interference (Min & Ichim, 2010) or by the specially designed RNA-binding proteins (Lunde et al., 2007; Filipovska & Rackham, 2012; Hall, 2016). The assembly/packaging signals look quite promising as the targets in the genomic ssRNA. The modified N proteins or their fragments can be used for similar purposes and may introduce defects in the nucleocapsid and make the virus less viable. The incorporation of assembly/packaging motifs into oligonucleotides immobilized on the surface of microarrays may facilitate the detection of coronaviruses by microarrays (for a review on microarrays see, e.g., Dufva, 2009).

*4.4. Comments on the specificity of motifs*

Presumably, the most working motifs (or, more exactly, the complete words defined in Section 2.4) participating in ssRNA-N proteins specific interactions are of 2–4 nt in length. They are frequent enough (see Table 2) and their coordinate positioning over the genome may provide specific cooperative interaction with N proteins. The close incorporation of the longer motifs would be too restrictive because of the protein coding function of the genomic ssRNA. However, the longer and rarer motifs may be multifunctional and may play the role of *cis*/*trans*-elements for the other molecular mechanisms during the virus life cycle. This conclusion looks nearly definite for the pairwise motifs at the step $s = 84$ such as ATTATAATTATAAAT (SARS-CoV; the start sites 22711 and 22795) and ATTATAATTA (isolates of SARS-CoV-2; sites 22766 and 22850; 22810 and 22894; 22757 and 22841, respectively). Note that the positions of these motifs on the genomes are also closely conserved. The same concerns the longest motifs found at $s = 54$ in the genome of SARS-CoV-2, TATTCAAACAATTGTTG (sites 3213, 3257, and 3204, respectively).

The specific binding of N proteins with ssRNA results in the lowering of free energy, which may approximately be assessed by the Boltzmann factor,

$$\varphi_b \propto \exp(-\Delta F/T); \; \Delta F = F_f - F_i < 0 \;, \tag{35}$$

Typically, the Boltzmann factor grows at the lower temperatures. This means that weakly specific effects should be more pronounced at the lower temperatures. Taking into account huge numbers of species in virus populations, even a small decrease in free energy may produce a significant impact and be advantageous for the natural selection.

## 5. Conclusion

The methods developed in this paper are quite general and can be applied to the detection of assembly/packaging signals in all viral genomes packaged within helical capsids including the other infectious coronaviruses such as 229E, NL63, OC43, HKU1, and MERS-CoV. The ssRNA genomes of numerous filamentous and rod-shaped plant viruses are also packaged within capsids with helical symmetry (Stubbs & Kendall, 2012; Solovyev & Makarov, 2016). As shown, combining NCF, DFT and DDFT provides efficient tools for the investigation of this problem. It is essential that dominating triplet periodicity $p = 3$ typical of protein coding regions in the viral genomes should be suppressed to discern the longer periodic patterns related to the assembly/packaging signals. After detection of periodic patterns and determination of their periods, the underlying motifs can be explicitly reconstructed by TAMGI. Generally, TAMGI can be efficiently used for data mining and search for *cis*/*trans*-elements in genomic sequences. The mutual experimental and bioinformatic analysis and the knowledge about the assembly/packaging mechanisms in viral genomes should facilitate the choice of the most efficient strategy in practical medical applications. The regular study of hidden quasi-periodic patterns is of basic interest for the virology.